\documentclass[12pt,preprint]{aastex}
\usepackage{natbib}
\usepackage{amsmath}
\def\hi{{{\rm H}\,{\sc i}~}}
\def\ovi{{{\rm O}\,{\sc vi}~}}
\def\ovii{{{\rm O}\,{\sc vii}~}}
\def\oviii{{{\rm O}\,{\sc viii}~}}
\def\oviin{{{\rm O}\,{\sc vii}}}

\def\civ{{{\rm C}\,{\sc iv}~}}
\def\siiv{{{\rm Si}\,{\sc iv}~}}
\def\chandra{{\it Chandra}~}
\def\chandran{{\it Chandra}}
\def\xmm{{\it XMM-Newton}~}                               
\def\suzaku{{\it Suzaku}~}

\def\gax{${_>\atop^{\sim}}$}
  
\begin{document}

\title{Probing the mass and anisotropy of the Milky~Way gaseous halo: Sight-lines
toward Mrk~421 and PKS2155-304}
\author{A. Gupta and S. Mathur\altaffilmark{1}}
\affil{Astronomy Department, Ohio State University, Columbus, OH 43210, USA}
\email{agupta@astronomy.ohio-state.edu}
\author{M. Galeazzi}
\affil{Physics Department, University of Miami, Coral Gables, FL 33124, USA}
\author{Y. Krongold}
\affil{Instituto de Astronomia, Universidad Nacional Autonoma de Mexico
Mexico City, (Mexico)}

\altaffiltext{1}{Center for Cosmology and
Astro-Particle Physics, The Ohio State University, Columbus, OH 43210}

\begin{abstract}

We recently found that the halo of the Milky Way contains a large
reservoir of warm-hot gas that contains a large fraction of the missing
baryons from the Galaxy. The average physical properties of this
circumgalactic medium (CGM) are determined by combining average
absorption and emission measurements along several extragalactic
sightlines. However, there is a wide distribution of both, the halo
emission measure and the \ovii column density, suggesting that the
Galactic warm-hot gaseous halo is anisotropic. We present {\it Suzaku}
observations of fields close to two sightlines along which we have
precise \ovii absorption measurements with \chandran. The column
densities along these two sightlines are similar within errors, but we
find that the emission measures are different:
$0.0025\pm0.0006$cm$^{-6}~$pc near the Mrk~421 direction and
$0.0042\pm0.0008~$cm$^{-6}~$pc close to the PKS2155-304 sightline.
Therefore the densities and pathlengths in the two directions must be
different, providing a suggestive evidence that the warm-hot gas in the
CGM of the Milky Way is not distributed uniformly. However, the formal
errors on derived parameters are too large to make such a claim.  In the
Mrk~421 direction we derive the density of $1.6^{+2.6}_{-0.8} \times
10^{-4}$ cm$^{-3}$ and pathlength of $334^{+685}_{-274}$ kpc.  In the
PKS2155-304 direction we measure the gas density of $3.6^{+4.5}_{-1.8}
\times 10^{-4}~$cm$^{-3}$ and path-length of $109^{+200}_{-82}~$kpc.
Thus the density and pathlength along these sightlines are consistent
with each other within errors. The average density and pathlength of the
two sightlines are similar to the global averages, so the halo mass is
still huge, over 10 billion solar masses.  With more such studies, we
will be able to better characterize the CGM anisotropy and measure its
mass more accurately. We can then compare the observational results with
theoretical models and investigate if/how the CGM structure is related
to the larger scale environment of the Milky Way.

We also show that the Galactic disk makes insignificant contribution to
the observed \ovii absorption; a similar conclusion was also reached by
Henley and Shelton (2013) about the emission measure. We further argue
that any density inhomogeneity in the warm-hot gas, be it from clumping,
from the disk, or from a non-constant density gradient, would strengthen
our result in that the Galactic halo path-length and the mass would
become larger than what we estimate here. As such, our results are
conservative and robust.

\end{abstract}

\section{Introduction}

Late-type galaxies in the Local Universe are missing more than 70\% of
their baryonic mass. A large fraction of these missing baryons are
likely embedded in the low density, warm-hot gas extended over 100 kpc
in the circumgalactic medium (CGM) or galactic halo \citep{Silk 1977,
White1978, Sommer2006}. Alternatively, the missing baryons from galaxies
may have escaped from the potential wells of galaxies but reside in
their parent groups or clusters \citep[e.g.,][]{Humphrey2011}. Finally,
the missing baryons can be in the form of local warm-hot intergalactic
filament \citep[WHIM, ][]{Cen1999,
Mathur2003,Nicastro2002,Nicastro2003}. Whether in the CGM or in their
group or cluster medium, these missing baryons are expected to be in a
warm-hot gas phase at temperatures between $10^{5}-10^{7}~$K.

The UV metal absorption lines through external galaxies provide strong
evidence for the presence of the warm gas ($T < 10^{6}$~K) in the CGM.
Recently \citet{Tumlinson2011} probed CGM of 42 galaxies with \ovi
absorption lines.  They found that the CGM around star-forming galaxies
is extended, to over 100 kpc and the mass content in the ionized gas is
over $10^{9}M_{\odot}$. But the mass of the gas probed by UV absorption
lines is significantly below what is required to account for the
galactic missing baryons; it falls short by an order of magnitude.

The warm-hot gas at temperatures $10^{6}-10^{7}~$K can only be detected
in the soft X-ray band, probed by even more highly ionized metals such
as \oviin. There are observational evidences that spiral galaxies have
reservoirs of ionized warm-hot gas in the CGM. Recently
\citet{Williams2013} investigated intervening X-ray absorption line
systems toward H2356-309 observed by \citet{Buote2009}, \citet{Fang2010}
and \citet{Zappacosta2012}. They found that three of the four absorption
systems originate within virial radii of nearby galaxies along the
sightline with projected distances of 100s of kpc, implying that these
intervening X-ray absorption lines probe CGMs of external galaxies
instead of the WHIM. \citet{Anderson2013} detected X-ray emission around
isolated galaxies observed in the ROSAT All-Sky Survey (RASS) extended
out to 50~kpc.  They estimate the median masses of the hot gas extending
out to 200~kpc around early-type and luminous galaxies of $1.5 \times
10^{10}~M_{\odot}$ and $3.3 \times 10^{10}~M_{\odot}$,
respectively. There are also other reports of detection of hot gaseous
halo around giant spiral galaxies such as NGC1961 \citep{Anderson2011}
and UGC12591 \citep{Dai2012} (see also Mathur et al. 2008). Though the
reported masses in warm-hot halos of these galaxies are significant,
they are not a major contributor to the galactic missing baryons. It is
possible that the reported masses are underestimated because studies
based only on emission are biased toward detecting only higher density
gas.

Milky Way offers an ideal opportunity to characterize the warm-hot CGM,
both in absorption and emission.  The UV absorbers provide strong
evidence for the presence of the warm gas ($T < 10^{6}$~K) through
absorption lines of \siiv, \civ \citep{Lehner2011} and \ovi (Sembach et
al. 2003). The hotter phase of the warm-hot gas, at temperatures
$10^{6}-10^{7}~$K, can again be probed by even more highly ionized
metals. Indeed, absorption lines due to \ovii and \oviii at redshift
zero have been detected in X-rays toward extragalactic sight-lines by
\chandra and \xmm \citep[hereafter Paper-I]{Fang2006,
Bregman2007,Williams2005, Gupta2012}. The presence of the hot halo
observed through $z = 0$ X-ray absorption lines is generally agreed
upon, but its density, path-length, and mass is a matter of debate
\citep{Mathur2012}.

In Paper-I, we performed a comprehensive survey of \ovii and \oviii
absorption lines at $z = 0$ using \chandra observations.  We found that
the \ovii absorption lines along several sightlines are saturated and
the implied column densities are larger than in previous studies. We
also found a large covering fraction ($72$\%) of the absorbing gas. We
used newer and better measurements of the emission measure to solve for
the density and the path-length of the absorbing/emitting plasma. With
these improvements we found that there is a huge reservoir of ionized
gas around the Milky Way, with a mass of $\geq6\times10^{10}~M_{\odot}$
and a radius of over $100~$kpc, consistent with theoretical models
\citep{Feldmann2013, Fang2012}. The mass probed by this warm-hot gas is
larger that in any other phase of the CGM and is comparable to the
entire baryonic mass of the Galactic disk of $6\times
10^{10}~M_{\odot}$.  The baryonic fraction $f_{b}$ of this warm-hot gas
varies from 0.09-0.23 depending on the estimates of the virial mass of
the Milky Way, from $10^{12}~M_{\odot}$ to $2.5 \times
10^{12}~M_{\odot}$ \citep{Boylan2013}, bracketing the Universal value of
$f_{b}=0.17$.

Until recently most of the absorption/emission studies of the warm-hot
gas in the Milky-way galactic halo assumed an average emission measure
(EM) over the sky, and the recent value used in Paper-I is
$EM\sim0.003$~cm$^{-6}$~pc. However, analysis of the RASS data
\citep{Snowden2000} and subsequent shadow observations have shown that
there is significant spatial variation in the Galactic halo emission
\citep[and references therein]{Gupta2009}.  A systematic study of the
emission measure of the hot gas in the Galactic halo was conducted
recently by \citet{Yoshino2009} and \citet{Henley2010} with the \suzaku
and \xmm archival data respectively. They observed that the halo
temperature is fairly constant across the sky ($\approx
1.8-2.4\times10^6~$K), whereas the halo emission measure varies by an
order of magnitude.  This may affect the estimates of density and
path-length of the halo warm-hot gas and the subsequent mass
measurement.

Instead of the averages of emission and absorption measurements used in
previous studies, using emission measures close to absorption sightlines
would be a significant step forward for probing the anisotropy of the
Galactic CGM.  With this goal, we re-examined the emission/absorption
measurements in directions of bright quasars Mrk~421 and PKS~2155-304 at
redshifts of 0.030 and 0.116, respectively. These extragalactic sources
have highest quality \chandra and \xmm grating spectra, which have been
analyzed in detail by several authors \citep[Paper-I]{Nicastro2002,
Nicastro2005, Kaastra2006, Williams2006, Williams2007, Fang2006,
Bregman2007}.  Both \chandra and \xmm detected the local (z=0) \ovii and
\oviii absorption lines towards these sightlines with high
confidence. In this work, we used the results from Paper-I for
absorption measurements, which are in agreement with other works.  For
the emission measurements we analyzed the \suzaku archival data.
\citet{Hagihara2010} have published the \suzaku data from adjacent
fields of PKS~2155-304, but we re-analyze these data for consistency
with the Mrk~421 field and also because we use a different method (as
described in \citet{Henley2010}) to constrain the Galactic halo emission
measure in this direction. The paper is organized as follows: in section
2 we present \suzaku observations, detailing the process for extracting
the Galactic halo emission measure. In section 3 we present spectral
modeling and in section 4 we review absorption measurements from Paper
I. In section 5 we present combined absorption/emission analysis and the
discussion is in \S 6. We conclude in \S 7.

\section{Suzaku Observations and Data Reduction}

\suzaku observed four X-ray blank sky fields adjacent to PKS~2155-304
and Mrk~421, two in each direction, in June 2008 and November 2009
respectively. Fig.~1 shows the RASS (0.1-2.4 keV) images in the vicinity
of PKS~2155-304 and Mrk~421, along with the \suzaku pointing of nearby
blank sky fields. The observation IDs, dates, pointing directions and 
exposure times are summarized in Table 1.

In this work we only used the data form X-ray Imaging Spectrometer (XIS)
on board \suzaku. For data reduction we followed the procedure described
in details in \citet{Gupta2009} and \citet{Mitsuishi2012}.  The data
reduction and analysis were carried out with HEAsoft version 6.13 and
XSPEC 12.7.0 with AtomDB ver.2.0.1.  In addition to standard data
processing (e.g. ELV $> 5$  and DYE ELV $> 20$), we performed a data
screening with the cut-off-rigidity (COR) of the Earth's magnetic field,
which varies as \suzaku traverses its orbit. During times with larger
COR values, fewer particles are able to penetrate the satellite and the
XIS detectors.  We excluded times when the COR was less than 8 GV, which
is greater than the default value (COR 4 GV) for all four observations,
as lowest possible background was desired.

We produced the redistribution matrix files (RMFs) by {\it xisrmfgen}
ftool, in which the degradation of energy resolution and its position
dependence are included. We also prepared ancillary response files
(ARFs) using {\it xissimarfgen} ftool \citep{Ishisaki2007}, which takes
into account the spatially varying contamination on the optical blocking
filters of XIS sensors which reduce the detector efficiency at low
energies \citep{Koyama2007}.  For the ARF calculations we assumed a
uniform source of radius $20^{\prime\prime}$ and used a detector mask
which removed the bad pixel regions.

Although reduced by the Earth's magnetic field, \suzaku still has a
noticeable particle background. We estimate the total instrumental
background from a database of the night Earth data with the {\it
xisnxbgen} ftools task \citep{Tawa2008}.  Night Earth data were
collected when the telescope was pointed at the night Earth (elevation
less than -5 degree, and pointed at night side rather than day). The
instrumental background was removed from the event files of the
observations.  The event files were also carefully screened for
telemetry saturation and other artifacts.

Determination of the Galactic halo emission measure is highly complex.
Until recently, ROSAT $3/4$ keV maps (Snowden et al. 1997) were used for
this purpose (e.g. in Bregman et al. 2007), but ROSAT had insufficient
spectral resolution to separate the foreground (Local Bubble (LB), plus
solar wind charge exchange, (SWCX)) and background (Galactic halo plus
extragalactic emission) components. Observations with \xmm and \suzaku
are much better for this purpose, with their large field of view and
spectroscopic capabilities. \citet{Gupta2009} and \citet{Henley2010} 
have discussed in details the process of extracting Galactic halo
emission measure, but the complexity of this analysis warrants some
discussion here. 

For determination of the Galactic halo emission measure, which is such a
weak signal, we are required to perform careful analysis with several
steps: (1) Determine the contribution from the LB and SWCX. We have
taken care to choose data when the variable component of the SWCX is
minimum (\S 2.1). Any residual and non-varying component of the SWCX is
determined by modeling the soft X-ray spectrum with an unabsorbed
thermal plasma emission model (\S 3.1). A detailed estimate of the
individual SWCX and LB contribution based on solar wind flux and ROSAT
All Sky Survey (RASS) data has also been used to verify the accuracy of
our approach and the systematics.  (2) Determine the contribution from
unresolved extragalactic background sources. This is modeled with an
absorbed power law (\S 3.2).  (3) Finally determine the hotter Galactic
halo contribution, modeled as an equilibrium thermal plasma component
absorbed by the gas in the galactic disk (\S 3.3).



\subsection{Reducing SWCX component}

Diffuse X-ray background at energies below 1~keV is severely affected by
SWCX, which varies both in spectral composition and flux on  scales of
hours to days. The SWCX is produced when highly ionized solar wind
particles interact with neutral gas, causing an electron jump from the
neutral atom to an excited level in the ion.  The electron then cascades
to the lower energy level of the ion, emitting soft X-rays and other
lines in the process. The neutral atoms may be in the outer reaches of
the Earth's atmosphere (giving rise to geocoronal emission), or they may
be in the interstellar material flowing through the solar system (giving
rise to heliospheric emission). The minimum SWCX plus LB contribution to
\ovii line intensities has been estimated to be about 2~Luminosity Units
(LU; ph~s$^{-1}$~cm$^{-2}$~sr$^{-1}$) \citep{Yoshino2009, Henley2013}.

As a part of an XMM-Newton ``Large Program'' we have also conducted a
monitoring campaign of the high latitude molecular cloud MBM12 to study
the temporal variation of the heliospheric SWCX emission.  By choosing a
molecular cloud we remove the non-local X-ray background and maximize
the SWCX signal. The campaign used 5 different pointings spanning a time
of almost 10 years and correlates the oxygen emission with solar wind
conditions (Galeazzi et al. 2013). The Suzaku data of Mrk~421 and
PKS2155-304 were obtained during a minimum of the Sun's 11-year cycle, a
period that is well characterized by the MBM12 campaign. The data
indicate that the heliospheric SWCX contribution to \ovii in the MBM12
direction during the Mrk~421 and PKS2155-304 observations was about 1
LU. 


While the SWCX contribution was low, it was non-zero and it could be
further minimized with proton flux filtering \citep{Smith2007,
Yoshino2009, Yoshitake2013}. This is because the geocoronal component of
the SWCX depends on the solar wind proton flux incident on the Earth and
on the sightline path through the magnetosheath. The higher proton flux
indicates higher ion flux that produces a contamination of specific
lines through the charge exchange process as described in
\citet{Ezoe2010}. We obtained the solar wind proton flux data from
OMNIWeb\footnote{ (http://omniweb.gsfc.nasa.gov/)}. Figures 2 \& 3 show
the lightcurves of Mrk~421 and PKS~2155-304 off-fields observations and
of solar wind proton flux during these exposures.  We removed the
portions of \suzaku observations that were taken when solar wind proton
flux exceeded $4\times10^{8}~$cm$^{-2}$~s$^{-1}$; this threshold is
typically used to reduce the geocoronal SWCX contribution
\citep{Mitsuda2007, Hagihara2010, Mitsuishi2012}. 

Even after applying the above proton flux filtering, the spectrum could
be contaminated by geocoronal SWCX, as this threshold is higher than the
average proton flux of $2.8\times10^{8}~$cm$^{-2}$~s$^{-1}$ at 1~AU
\citep{Henley2010}. However, if we apply tighter constraints on proton
flux filter we loose large portions of the \suzaku exposure time, so the
residual SWCX contribution to the soft X-ray flux is taken into account
with spectral modeling (\S 3.1).

\section{Spectral Modeling}
 
We cleaned XIS1 spectra extracted from each \suzaku observation
following the above procedure and then performed spectral modeling. As
noted above, the spectra have three distinct components: (1) a foreground
component made of residual SWCX and LB; (2) a background 
component made of unresolved extragalactic sources; and (3) the Galactic
halo component. Below we discuss each of these in turn.

\subsection{Shape and normalization of the foreground component.}

We model the foreground as an unabsorbed plasma in collisional ionization
equilibrium (CIE) with thermal emission. The parameters of this model
are plasma temperature and normalization.  We fixed the temperature $T
=1.2 \times 10^{6}~$ K as suggested by data from RASS
\citep{Snowden1998, Snowden2000, Kuntz2000} and further supported by
current X-ray shadow observations \citep[for summary see][]{Gupta2009}.

To determine the normalization we used \citet{Snowden2000} catalog of
SXRB shadows, consisting of 378 shadows in the RASS. This catalog
contains the foreground and background R12 (1/4 keV)
count-rates. Following \citet{Henley2010}, we found 5 shadows in the
catalog closest to our sightlines, and averaged their foreground
count-rates, weighted by the inverse of their distances from the \suzaku
sightlines:

\begin{equation}
Foreground~ R12~countrate=\frac{\sum R_{i}/ \theta_{i}}{\sum 1/ \theta_{i}}
\end{equation}

where $R_{i}$ is the foreground R12 count-rate for the $i$th shadow, and
$\theta_{i}$ is the angle between the center of the $i$th shadow and the
\suzaku sightline. Since foreground intensity varies fairly mildly over
the sky \citep{Snowden2000}, this method provides a reasonable estimate
of the foreground intensity at high latitudes ($l > 30\deg$) in the R12
band. Then we extrapolated the foreground spectrum from the R12 band to
the $0.5-1.0~keV$ soft X-ray band.

In the adjacent fields of Mrk~421 and PKS~2155-304, we calculated the
average foreground R12 count-rate of (767$\pm$243)$\times
10^{-6}$~counts~s$^{-1}$~arcmin$^{-2}$ and (397$\pm$143)$\times
10^{-6}$~counts~s$^{-1}$~arcmin$^{-2}$ respectively.  Using the above
calculated count-rate, we determine the normalization of the foreground
thermal component assuming $T =1.2 \times 10^{6}~$ K.

\subsection{Background component.}

The background component from unresolved extragalactic sources was
modeled with an absorbed power law. The parameters of the model are the
Galactic column density, which was held fixed, and power-law slope and
normalization, which were left as free parameters in the spectral fit;
the results are given in Table 2.  We point out that the power low
component is the dominant contributor to the spectrum above 1.5 keV, but
does not dominate at softer energies where the Galactic halo component is
present.

\subsection{Galactic halo component.}

The Galactic halo emission was modeled as an equilibrium thermal plasma
component absorbed by the cold gas in the Galactic disk. The parameters of
this model are the plasma temperature, normalization, and the Galactic
column density. For the absorption, we used the XSPEC {\it wabs} model,
which uses cross-sections from Wisconsin \citep{Morrison1983} and uses
the \citet{Anders1989} relative abundances; the column density along our
sightlines was fixed to the observed values \citep{Dickey1990}. For
plasma thermal emission, we used the Astrophysical Plasma Emission Code
({\it APEC}: Smith \& Cox 2001) and the temperature and normalization
were left as free parameters of the fit.

Finally, \suzaku spectra of Mrk~421 and PKS~2155-304 off-fields were
fitted using all the three components noted above.  We measured the
galactic halo temperature of $\log T = (6.31\pm0.04)$ \& $(6.32\pm0.02)~
$K and emission measure of $ (2.4\pm0.5) \times 10^{-3}$ \& $(2.6\pm0.4)
\times 10^{-3}~$cm$^{-6}~$pc near Mrk~421 off-field~1 and off-field~2
respectively. 


The galactic halo emission measure and temperature near PKS~2155-304
off-field~1 and off-field~2 are $ (4.0\pm0.5) \times 10^{-3}$ \&
$(4.3\pm0.4) \times 10^{-3}~$cm$^{-6}~$pc and $\log T = (6.37\pm0.02)$
\& $(6.35\pm0.02)~ $K respectively. 


The temperature and emission measure of Galactic halo thermal component
were consistent between the two fields near PKS~2155-304 and Mrk~421
sightlines. Thus to obtain better constraints we fitted the above
mentioned models simultaneously to both the off-field spectra with a
single set of parameters (except for the power-law) in each
direction. The fits are shown in Fig. 4, and the model parameters are
presented in Table 2.

\subsubsection{Additional contamination due to SWCX.}

From each spectrum, we also evaluated line centers and surface
brightness of \ovii K$\alpha$ and \oviii K$\alpha$ emission lines.  In
the best fit model of each observation, we replaced the {\it APEC} model
with {\it VAPEC} model for the Galactic halo component, whose oxygen
abundance was set to $0$ and added two Gaussian lines for \ovii
K$\alpha$ and \oviii K$\alpha$ emission. The abundances of other
elements were fixed to be 1 solar, and temperatures of {\it VAPEC}
components were fixed to the best-fit values for the individual
fits. The line centers and the surface brightnesses were derived from
these spectral fits.

The \ovii intensities estimated from spectral fitting were consistent
within 1~sigma error between Mrk~421 off-field~1 and off-field~2
observations (Table 3). The PKS~2155-304 off-field~1 and off-field~2
intensities, however, were different, 
$7.4\pm0.7~$ph~s$^{-1}~$cm$^{-2}~$sr$^{-1}~$(LU) and $8.3\pm0.7~$LU
respectively. We investigated whether this difference is due to
contamination from geocoronal SWCX as follows. While during the Mrk~421 
pointings the solar wind proton flux was consistently below 
$2.8\times10^{8}~$cm$^{-2}$~s$^{-1}$, during the PKS~2155-304
off-field~1 observation, the proton flux varied considerably from less
than about $2.0\times10^{8}~$cm$^{-2}$~s$^{-1}$ to about
$4.0\times10^{8}~$cm$^{-2}$~s$^{-1}$ (figure 4). We extracted and
modeled the spectra during these intervals and found that the \ovii
intensity changed from $5.3\pm1.2$ to $7.8\pm1.1~$LU. During the
PKS~2155-304 off-field~2 observation the proton flux was always higher
than $2.0\times10^{8}~$cm$^{-2}$~s$^{-1}$ and the \ovii intensity
remained on higher side at $8.3\pm0.7~$LU. Considering the observational 
geometry for our pointings, we expect the contribution from geocoronal 
SWCX to the \ovii line to be negligible for proton fluxes below 
$2.0\times10^{8}~$cm$^{-2}$~s$^{-1}$. This is not true, however, for the 
higher flux encountered during the PKS 2155-304 pointings. 
Based on the numbers reported above, we estimate that the
contamination due to geocoronal SWCX is of the order of $\sim 2~$LU. We
therefor added this SWCX contribution to the normalization of the
foreground component and fitted the PKS~2155-304 off-field spectra
again. The parameters in Table 2 refer to this final fit.


\subsection{Systematic error on Emission measurement}

In our spectral model we fixed the temperature and normalization of the
foreground component, otherwise there would be degeneracy between the
foreground and Galactic halo intensities. This results in systematic
errors on Galactic halo emission measure due to our assumed spectrum of
the foreground component.  We have fixed the foreground normalization by
extrapolating the foreground spectrum from the R12 ($0.1-0.284~$keV)
band into the \suzaku $0.5-1.0$ keV band. However, this extrapolation
may not be a correct estimate of the foreground normalization, because
the spectral shapes in the two bands may be different. Additionally,
relative contributions of LB and SWCX emission are likely to differ in
these bands \citep{Kou2009a, Kou2009b}.

A conservative way to estimate the systematic error is to determine the
upper limit on the galactic halo emission measure.  To do that, we
estimated the lower limit on the combined emission of LB plus
heliospheric SWCX (the contribution from geocoronal SWCX has already
been discussed in details on section 3.3.1). As our observations were
performed during a time of low solar wind activity, a conservative
estimate is to assume that the minimum heliospheric SWCX contribution is
zero. For the lower limit on LB emission, we followed the procedure in
Lallement (2004). We found that the lower limit on the foreground
emission is $0.0029\pm 0.0004$ cm$^{-6}$pc resulting in the upper limit
on Galactic halo emission of $0.0003\pm 0.0005$ cm$^{-6}$pc and
$0.0051\pm 0.0004$ cm$^{-6}$pc in the directions of Mrk 421 and PKS
$2155-304$ respectively.  Thus we estimate the systematic errors on the
emission measure to be $0.0005$~cm$^{-6}~$pc and $0.0009$~cm$^{-6}~$pc
for the two directions.

We also used a method described in Henley et al. to estimate the
systematic errors by reanalyzing each spectrum, using the median
foreground R12 intensity (600$\times 10^{-6}$
counts~s$^{-1}$~arcmin$^{-2}$) to fix the foreground normalization
instead of that from the 5 nearby shadows (\S 3.1). The systematic
errors we get using this method are $0.0004$~cm$^{-6}~$pc and
$0.0005$~cm$^{-6}~$pc for the two directions. These are lower than our
previous estimates, so to be conservative we adopted errors estimated
above. Note that if the Galactic halo emission measure is systematically
lower, then the implied pathlength is higher, and the mass of the CGM
would also be higher (see Gupta et al. 2012, Mathur 2012).


\section{\chandra Absorption Measurements}

In Paper-I, we measured the equivalent widths (EWs) of $z=0$ \ovii
$K\alpha$ absorption line of $11.6\pm1.6~$m\AA~ and $9.4\pm1.1~$m\AA~
towards PKS~2155-304 and Mrk~421 respectively. We also detected the
$z=0$ \ovii K$\beta$ absorption toward both sight-lines with EWs of
$4.2\pm1.3~$m\AA~ and $4.6\pm0.7~$m\AA~ respectively. The observed
$\frac{EW(O_{VII}~ K\beta)}{EW(O_{VII}~ K\alpha)} $ ratio of
$0.36\pm0.12$ and $0.49\pm0.09$ (in comparison to expected ratio of
0.156), clearly shows that \ovii K$\alpha$ lines are saturated.  Thus
using the ``curve-of-growth'' analysis we constrained the \ovii column
densities of $\log (N_{OVII})=16.09\pm0.19$ and $16.22\pm0.23~$cm$^{-2}$
towards PKS~2155-304 and Mrk~421 respectively.  For detailed analysis of
absorption measurements, we refer the readers to Paper-I.

In Paper-I we also reported on the $z=0$ \oviii absorption lines in the
spectra of Mrk~421 and PKS2155-304. With column density ratio
of \ovii to \oviii, we constrained the temperature of the absorbing gas
in these sightlines to $\log T=6.16\pm 0.08~$K and $\log T=6.27\pm
0.05~$K respectively. These estimates are comparable to the temperature
obtained from galactic halo emission model (Table 2).  Given that \ovii
absorption and emission trace warm-hot gas at similar temperatures, it
is reasonable to assume that they arise in the same gas (but see \S 6.2
and \S 6.3 where we relax this assumption). We can then combine these
two diagnostics to derive physical properties of the warm-hot gas.

\section{Combined Emission/Absorption Analysis}

The absorption lines measure the column density of gas $N_H= \mu n_e L$, 
where $\mu$ is the mean molecular weight $\approx 0.8$, $n_e$ is 
the electron density and $L$ is the path-length.  The emission measure, 
on the other hand, is sensitive to the square of the number density of 
the gas ($EM= n_e^2 L$, assuming a constant density plasma).  Therefore 
a combination of absorption and emission measurements naturally 
provides constraints on the density and the path-length of the 
absorbing/emitting plasma.

\subsection{Mrk~421}

The CGM \ovii column density towards Mrk~421 is $16.22\pm0.23~$cm$^{-2}$
and the average emission measure of two fields 30$^{\prime}$ away on
opposite sides of the Mrk~421 sightline is
$0.0025\pm0.0003\pm0.0005$~cm$^{-6}$~pc. Combining absorption and
emission measurements, we solve for the electron density and path-length
of \ovii emitting/absorbing plasma, following the procedure in
Paper-I. Towards the sight-line of Mrk~421 the electron density is given
by:

\begin{equation}
n_e= (1.6^{+2.6}_{-0.8} \times 10^{-4})~ (\frac{0.5}{f_{O VII}})^{-1} cm^{-3}
\end{equation}

and the pathlength is: 

\begin{equation}
L = (334^{+685}_{-274}) (\frac{8.51 \times 10^{-4}}{(A_O/A_H)}) 
(\frac{0.5}{f_{O VII}})^2 (\frac{0.3Z_{\odot}}{Z}) ~kpc
\end{equation}

where the solar oxygen abundance of $A_{O}/A_{H}=8.51 \times 10^{-4}$ 
is from \citet{Anders1989}, $f_{OVII}$ is the ionization fraction of 
\ovii, and $Z=0.3Z_{\odot}$ is the metallicity.

\subsection{PKS~2155-304}

Toward the sight-line of PKS~2155-304, the halo \ovii column 
is $16.09\pm0.19$cm$^{-2}$ and the emission measure is 
$0.0042\pm0.0003\pm0.0007$~cm$^{-6}$~pc. Combining absorption and 
emission, we determine the electron density of

\begin{equation}
n_e= (3.6^{+4.5}_{-1.8} \times 10^{-4}) (\frac{0.5}{f_{O VII}})^{-1} cm^{-3}
\end{equation}

and the pathlength of: 

\begin{equation}
L = (109^{+200}_{-82}) (\frac{8.51 \times 10^{-4}}{(A_O/A_H)}) 
(\frac{0.5}{f_{O VII}})^2 (\frac{0.3Z_{\odot}}{Z}) ~kpc
\end{equation}

We note that the large error bars on the gas density and path-length 
are dominated by systematic uncertainties in the \ovii column density 
measurements due to line saturation. These error bars do not represent 
the statistical limit of the measurement. The \ovii absorption line EWs and 
galactic halo emission measure are determined with more than 5$\sigma$ 
significance.


\section{Discussion}

\subsection{Comparison with Paper-I}

In Paper-I we measured the average \ovii column density of $\log
N(O_{VII})=16.19\pm0.08~$cm$^{-2}$. In Mrk~~421 and PKS~2155-304
sightlines the measured \ovii column densities are $\log N_{O
VII}=16.22\pm0.23$ \& $16.09\pm 0.19~$cm$^{-2}$, respectively. The best
fit value toward Mrk~421 is somewhat larger than the average and that
toward PKS~2155-304 is somewhat smaller than the average (Fig. 5), but
they are consistent with the average and with each other, within errors.

In Paper-I we also assumed the average emission measure of $0.0030\pm
0.0006~$cm$^{-6}$~pc. The emission measures towards Mrk~421 and
PKS~2155-304 of $0.0025\pm0.0006$ \& $0.0042\pm0.0008~$cm$^{-6}~$pc
respectively, are significantly different from each other (Fig 5). 

Thus in the two sightlines that we consider, the observed column
densities are similar, but observed emission measures are different, so
their densities and pathlengths must be different.  This is because the
gas density and path-length are proportional to $\frac{EM}{N_{OVII}}$
and $\frac{N_{O VII}^{2}}{EM}$ respectively. Thus in the direction
toward Mrk~~421, lower emission measure brings the gas density lower
while path-length higher than average. The opposite is true for the
PKS2155-304 direction. 


Towards Mrk~421 we calculated the density and the path-length of
warm-hot gas of $1.6^{+2.6}_{-0.8} \times 10^{-4}~$cm$^{-3}$ and
$334^{+685}_{-274}$ kpc respectively.  On the other hand, near
PKS2155-304 sight-line, we measured the gas density of
$3.6^{+4.5}_{-1.8} \times 10^{-4}~$cm$^{-3}$ and path-length of
$109^{+200}_{-82}$ kpc. Thus the density and pathlength in the two
directions are similar within (large) errors.

This investigation on comparing absorption and emission measurements
from adjacent fields provides suggestive evidence that the warm-hot gas
in the CGM of the Milky Way is not distributed uniformly. However, the
formal errors on derived parameters are too large to make such a claim.
We need to perform similar studies for a number of sight-lines to probe
the anisotropy of the MW CGM.  The mean density and path-length from
these two sightlines, of $\sim 2.6 \times 10^{-4}~$cm$^{-3}$ and $\sim
221~$kpc, are consistent with the average from the entire \chandra
sample (Paper-I), so the mass measurement is unlikely to be very
different.

\subsection{Galactic disk contribution to observed absorption/ emission?}

Our interest is to determine the physical properties of the warm-hot gas
in the Galactic halo. For this purpose, in Paper-I we chose our targets
to be of high Galactic latitude $|b|$\gax $20^{\circ}$. The two targets
in this paper, Mrk~421 and PKS$2155-304$ have $b= 65.03$ and $b=
-52.245$ degrees respectively, so the observed \ovii absorption should
have minimal contribution from the Galactic thick disk. In order to
investigate further whether the observed \ovii absorption is dominated
by the disk of the Galaxy, we looked for the dependence on the observed
\ovii equivalent width (EW) on $\sin(|b|)$; we expect a clear
anticorrelation between these two parameters if the absorption is
dominated by the disk. In figure 6 we have plotted EW vs. $\sin(|b|)$
for all 21 sightlines in Paper I in which \ovii absorption was
detected. We see that the data do not show significant anticorrelation
(Spearman rank correlation coefficient $\rho=-0.24$, with the chance
probability of $0.14$).

The real anticorrelation we expect, however, is between $\sin(|b|)$ and
column density, not with EW, but we do not have column density
measurements for all the 21 sightlines. Given that saturation
effects are important in cases of high column densities, there may be
crowding at high EW values. Therefore, if the disk contribution is
significant we should find that the distribution of EWs is different in
low- and high- values of $\sin(|b|)$. We tested this using the Student's
t-test and found that $t=0.937$, with the probability that the
differences are due to chance $=0.361$. Thus once again we do not see
any significant differences with $\sin(|b|)$.  This suggests that the
Galactic disk does not contribute significantly to the observed \ovii
absorption.

Henley \& Shelton (2013) also came to the same conclusion for emission
 measure. They studied 110 \xmm sightlines and detected emission from
 the million degree gas in about $80\%$ of them. They find that the high
 latitude sky has a patchy distribution of warm-hot gas, but they find
 no evidence for the observed emission measure to decrease with
 increasing Galactic latitude (see their figure 5). Therefor they argue
 that the emitting plasma does not have a disk-like geometry.

Thus, both absorption and emission measurements show no evidence for a
significant disk contribution. Nonetheless, if there is {\it any}
contribution from the disk, it would end up strengthening our result in
that it would make the inferred halo path-length larger and so the mass;
this is explained in detail below.

\subsection{Assumption of a constant density profile.}

In deriving the Galactic halo pathlength and density, we have assumed
the density to be constant, as we did in Paper I. A realistic halo is
likely to have some density profile that falls with radius, so it is
imperative to investigate how reasonable our assumption is and what are
the effects of our assumption on our results. This is particularly
important because the EM is biased toward high density plasma, while the
absorption is sensitive to integrated density.

 The observations of other galaxies support the assumption of an extended 
density profile. As noted in \S 1, \citet{Williams2013}
investigated intervening X-ray absorption line systems toward
H$2356-309$ observed by Buote et al. (2009), Fang et al. (2010) and
Zappacosta et al. (2010). They found that three of the four absorption
systems originate within virial radii of nearby galaxies or groups with
projected distances of 100s of kpc. These observations give evidence for
extended warm-hot halos around other galaxies. The $z=0.030$ system in
Williams et al. is particularly relevant for the present discussion
because the observed sightline passes through the halo of a nearby
galaxy. The observed column density of this absorption system is $\log
N_{OVII}=16.8^{+1.3}_{-0.9}$ at an impact parameter of D$=90$ kpc from a
nearby galaxy with virial radius of R$=160$ kpc. The path-length of the
absorber is then $2 \sqrt{R^{2}-D^{2}}=264.6$ kpc. From the path-length
and the column density we calculate the density =$7.4\times 10^{-4}
cm^{-3}$ (for \ovii ionization fraction and metallicity as in
Paper-I). This shows that such a high density, even more than what we
calculated for the MW halo, is present out to about a hundred kpc from
another galaxy as well. This not only shows a MW-type halo around
another galaxy, it also shows that the assumption of a flat density
profile is reasonable. There is also another circumstantial evidence
that the Milky Way warm-hot halo has an extended gas
profile. \citet{Grcevich2009} explain the non-detection of \hi in most
dwarf spheroidal galaxies near the Milky Way by ram pressure stripping
by low density ($\sim 10^{-4}~$cm$^{-3}$) hot gas extended out to $\geq
70~$kpc. This again suggests that our assumption of a single
temperature, constant density, extended plasma is reasonable (see also
Mathur 2012).

Some authors have used different models to describe density gradients
such as a simple exponential profile or a $\beta$-model (e.g. Mathur et
al. 2008 and references therein). In the simulations of
\citet{Feldmann2013}, the density is roughly constant above the Galactic
disk out to about 100 kpc. \citet{Fang2012} considered a hot halo that
is distributed as adiabatic gas with a polytropic index of $5/3$, in
hydrostatic equilibrium within the Milky Way's dark matter halo with NFW
profile (Navarro, Frenk, \& White 1997). They also consider a thick disk
profile and a cuspy NFW profile and compare these models with the
observations of dwarf satellite galaxies (Grcevich \& Putman 2009). They
find that a Maller \& Bullock type flat, extended profile of the hot gas
is preferred (see their figure 1). While this is not exactly a constant
density profile, it is flat enough that constant density is a reasonable
approximation.

Thus observations of external galaxies as well as theoretical models
support a reasonably flat and extended (over 100 kpc) density
distribution and rule out thick-disk type profile for the hot halo
gas. Nonetheless, let us investigate the effect of a density gradient on
our results. Combining the information of EM with absorption column
density we measure the density and pathlength of the absorbing/ emitting
plasma, assuming a single density. Therefore we have density $n\propto
EM/N_H$ and so the path-length $L \propto N_H^2/EM$. Now let us say that
the density is actually not constant, but has a high density component
n(High) and a low density component n(Low) such that
$n(High)>>n>>n(Low)$. We are considering just two components here for
simplicity, but the argument can be generalized to many components or to
any density profile. The observed EM is then a sum of the EM from the
two components: $EM= n(High)^2L(High) + n(Low)^2L(Low)$ where L(High)
and L(Low) are the pathlengths of the high and low density components
respectively. Similarly, the observed column density $N_H =
n(High)L(High) + n(Low)L(Low)$ (ignoring constants).  To satisfy these
constraints, $n(High)>>n>>n(Low)$ necessarily implies that
$L(High)<<L<<L(Low)$. Thus the path-length of the low density component
L(Low) is necessarily larger than the pathlength calculated assuming a
single density. Given that the implied mass goes as $L^3$, the mass also
becomes more than that calculated assuming a single density. Thus we see
that {\it any} density gradient would make the Galactic halo more
extended and more massive than what we find under a single density
assumption. And this is true for any contribution from denser components
such as a thick disk or from clumping. This conclusion is mathematically
robust and can be generalized to any density gradient. Since this
argument may not be obvious, we have discussed it further in the
Appendix.

\subsection{Comparison with other works}

\subsubsection{PKS$2155-304$}

\citet{Hagihara2010} have compared the \ovii absorption with the
galactic halo emission in the adjacent field of PKS2155-304. They
measured the \ovii column density of
$15.76^{+0.09}_{-0.12}~$cm$^{-2}$\footnote {The reported errors from
Hagihara et al. are at 90\% confidence level}, which is smaller than our
measured value of $16.09\pm 0.19~$cm$^{-2}$.  They measured the \ovii
equivalent width of $13.3\pm2.8~$m\AA~ and the Doppler parameter {\it b}
of $294^{+149}_{-220}~$km~s$^{-1}$. Though their measured equivalent
width is consistent with our measurement, the Doppler parameter {\it b}
is much larger than our 1$\sigma$ constraints of
$35-94~$km~s$^{-1}$. Due to larger Doppler parameter they underestimate
the \ovii column density by a factor $\approx 2$. For the gas at
temperature of about $2 \times 10^{6}~$K, the velocity dispersion
parameter {\it b} should be of the order of 45~km~s$^{-1}$, consistent
with our measurement. While there could be additional contribution to
the line width due to microturbulence, $294$ km~s$^{-1}$ appears to be
excessively high.

Hagihara et al. estimated the lower and upper limits on galactic halo emission 
measure of $2.0\times 10^{-3}$ and $4.9 \times 10^{-3}~$cm$^{-6}$~pc, assuming 
3.5~LU for the foreground \ovii emission and zero normalization of foreground 
component, respectively. Our measured value of $0.0042\pm0.0006~$cm$^{-6}~$pc 
is within this range. 

Combining the absorption and emission measurements they determined the
halo path-length and density of $4.0^{+1.9}_{-1.4}~$kpc and
$(7.7^{+2.3}_{-1.7}) \times 10^{-4}~$cm$^{-3}$ respectively, assuming
solar metallicity and \ovii ionic fraction of 1. Hagihara et al. upper
limit on galactic halo path-length of $5.4~$kpc is in contradiction with
our measurement of $L>30~$kpc. This discrepancy is due to (a) \ovii
column density (their column is $\sim 2$ times smaller than ours), (b)
oxygen abundance ($Z_{\odot}$ versus 0.3$Z_{\odot}$), and (c) ${f_{O VII}}$
of $1$ versus $0.5$.

Cosmological simulations \citep{Toft2002, Sommer2006} of formation and
evolution of disk galaxies show that outside the galactic disk the mean
metallicity of gas is $Z = 0.2\pm0.1 Z_{\odot}$.  These values of
metallicities are also consistent with observational results for the
spiral galaxy NGC~891 \citep{Hodges2013}, the outskirts of groups
\citep{Rasmussen2009} and clusters of galaxies
\citep[e.g.,][]{Tamura2004}. Thus it is highly unlikely that outside the
galactic disk metallicity is as high as solar.  The lower value of \ovii
column density and higher metallicity used by Hagihara et
al. underestimate the pathlength by a factor of $\sim 13$. The
difference in ${f_{O VII}}$ would further add a factor of four (see
Paper-I for justification of ${f_{O VII}}=0.5$).
 
The other major difference between our analysis and Hagihara et al. is
of assumption of gas density distribution. Hagihara et al. favor the
exponential disk model in which temperature and density of the hot gas
falls off exponentially along the vertical direction. Such a profile is
based on Yao et al. (2007, 2009), required to account for the fact that
they measured different halo temperatures from emission and absorption
analysis. We do not have this requirement; we found similar temperature
for the Galactic halo from both absorption and emission analysis. In any 
case, see \S 6.3 above.

\subsubsection{Mrk~421}

\citet{Sakai2012} combine the absorption and emission measurements 
towards Mrk~~421 and measured the galactic halo path-length and density of 
$1.6^{+3.6}_{-0.9}~$kpc and $(1.5^{+0.7}_{-0.5}) \times 10^{-4}~$cm$^{-3}$ 
respectively. The reported path-length again is in contradiction with 
our measurements. Since this is a conference proceeding paper, 
we do not have detailed information about their analysis.

\subsubsection{Henley \& Shelton 2013}

Recently Henley \& Shelton (2013) published new results on the galactic
halo emission measure towards 110 \xmm sightlines. They found galactic
halo temperature is fairly uniform (median $= 2.22 \times 10^{6}~$K),
consistent with their previous analysis (Henley et al. 2010). However
their current estimates on galactic halo emission measure are
systematically lower, with median value of $1.9 \times
10^{-3}~$cm$^{-6}$pc compared to $3.0 \times 10^{-3}~$cm$^{-6}$pc from
the previous paper. As the gas density and path-length of galactic halo
are proportional to $\frac{EM}{N_{OVII}}$ and $\frac{N_{O VII}^{2}}{EM}$
respectively (\S 6.1), the lower emission measure increases the
path-length and decreases the density by a factor of $\sim 1.6$ from the
Paper-I values.  Thus mean path-length and mass of the hot gas in CGM of
the Milky-way becomes $L=382\pm160~$kpc and $M_{total}=(5.9\pm5.4)
\times 10^{11}~M_{\odot}$ (for $Z=0.3Z_{\odot}$), respectively, compared
to $L=239\pm100~$kpc and $M_{total}=(2.3\pm2.1) \times
10^{11}~M_{\odot}$ in Paper I. Thus, if this new value of the average
emission measure is correct, most of the Galactic missing baryons would
be accounted for by the warm-hot halo.

\section{Conclusions}

In paper-I we compared average absorption and emission measurements of the
 warm-hot gas in the CGM to derive its average physical properties such as
 density, pathlength and mass. The absorption column density as well as
 the emission measure vary considerably across the sky, so it is far
 more appropriate to determine emission measures close to absorption
 sightlines and derive physical properties along each sightline
 independently. In this paper we expand on the previous work by
 comparing absorption and emission along two sightlines: toward Mrk~~421
 and PKS2155-304. We present {\it Suzaku} observations of fields close
 to these sightlines. We find that along the Mrk~421 direction, the
 emission measure is lower than the average, while the opposite is true
 in the PKS2155-304 sightline.  The mean values from these two
 sightlines are consistent with the average, so the mass of the CGM is
 unlikely to be very different from that in Paper-I (but see \S 6.4.3).

In the two sightlines we consider, the observed column
densities are similar, but observed emission measures are different, so
their densities and pathlengths must be different. This provides a 
suggestive evidence that the warm-hot gas in the CGM of the Milky Way is
not distributed uniformly, but the formal errors on derived parameters
are too large to make such a claim.  In future studies we will explore
the entire distributions of column densities and emission measures
(fig. 6 ) for combined absorption and emission analysis along multiple
sightlines. Large fields of view and spectroscopic capabilities of \xmm
and {\it Suzaku} are ideal for determining emission measures close to
bright AGN sightlines, which are used for absorption measurements. With
more such studies, we will be able to better characterize the CGM
anisotropy and measure its mass more accurately. Moreover, we can then
compare the observational results with theoretical models (\S 6.3) and
investigate if/how the CGM structure is related to the larger scale
environment of the Milky Way.

\clearpage

\begin{figure}
\begin{center}
\includegraphics[width=6.5cm,height=6cm]{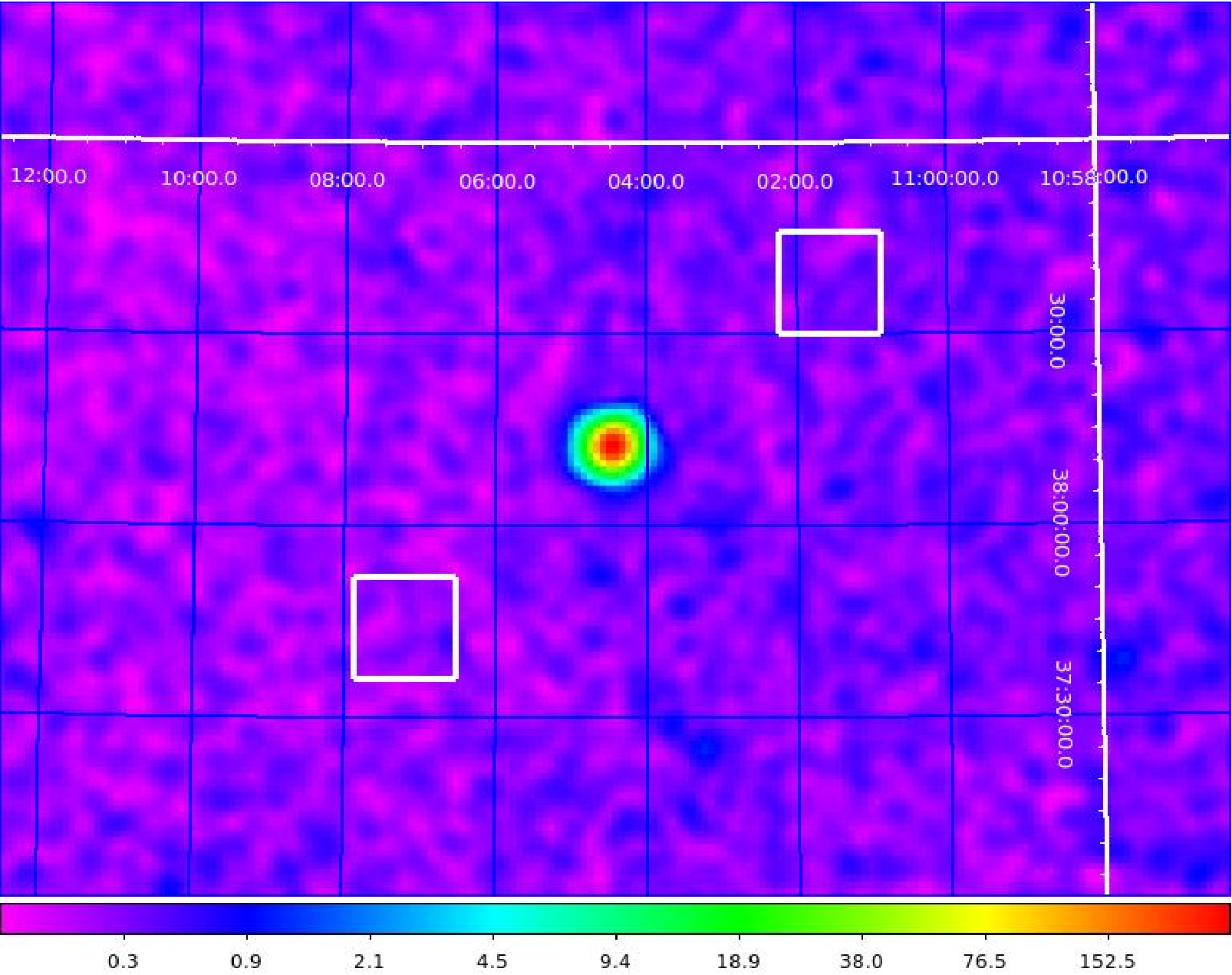}
\hspace{0.5in}
\includegraphics[width=6.5cm,height=6cm]{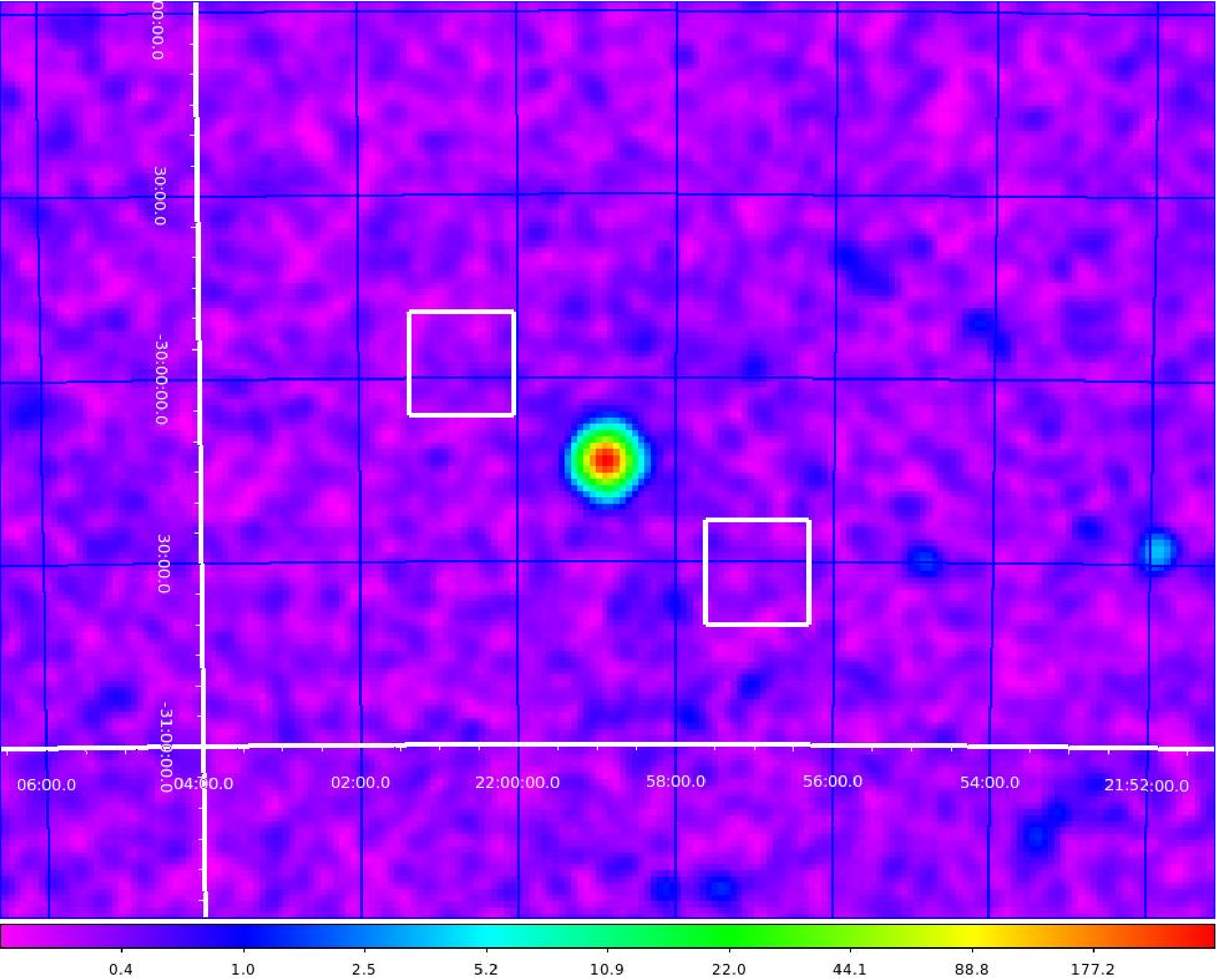}
\end{center}
\caption{RASS map ($4\deg \times 4\deg$) in the vicinity of Mrk~421 
({\it left}) and PKS~2155-304 ({\it Right}). The white squares show the 
Suzaku pointings used in this investigation, separated by $\sim 1.4 \deg$ 
and $\sim 1.0 \deg$ in case of Mrk~~421 and PKS2155-304, respectively. 
The gray scale represent the log of ROSAT broad band counts per pixel.
}
\label{fig1}
\end{figure}

\clearpage



\begin{figure}
\begin{center}
\includegraphics[width=10cm,height=10cm]{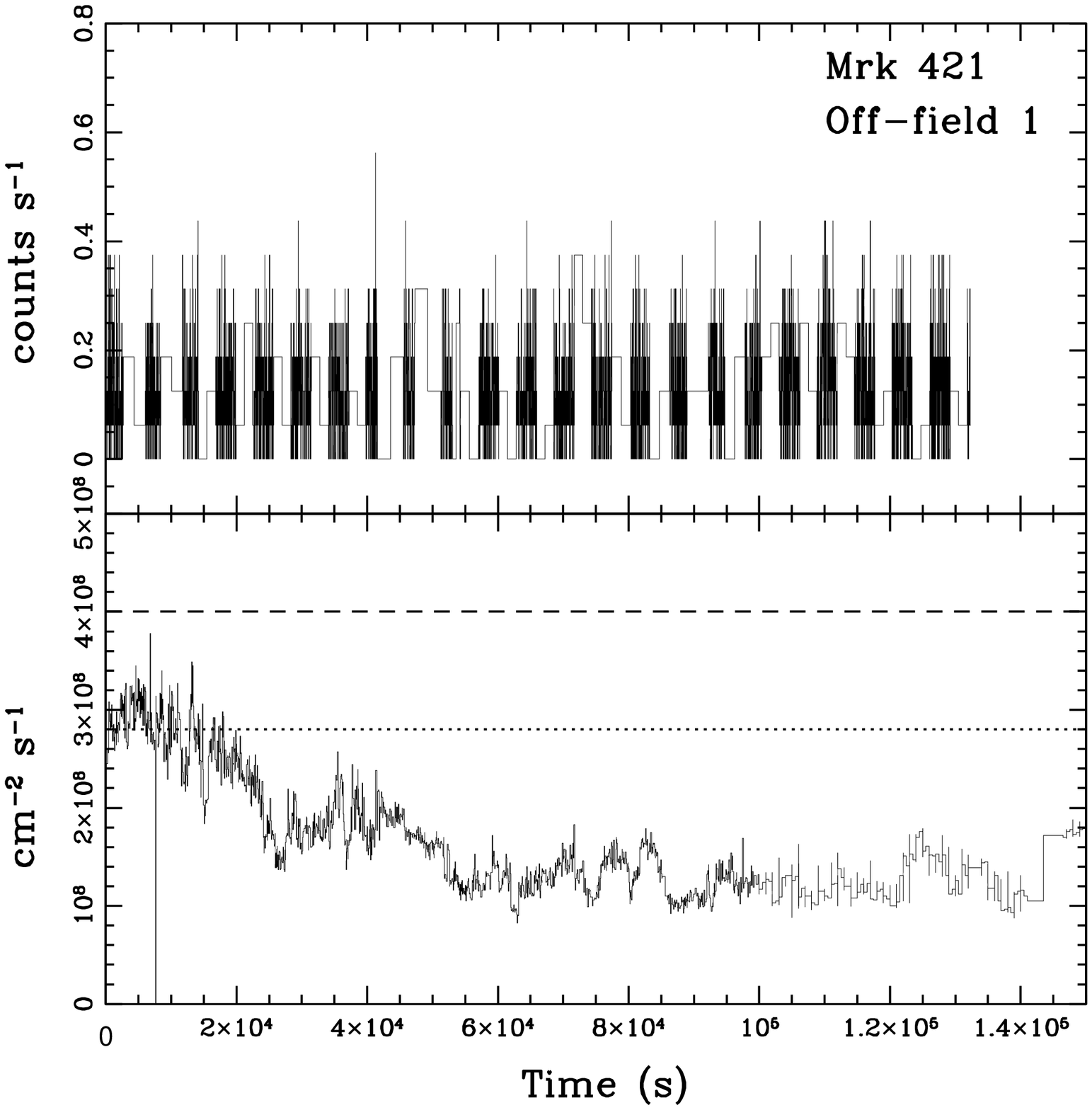}
\hspace{1cm}
\includegraphics[width=10cm,height=10cm]{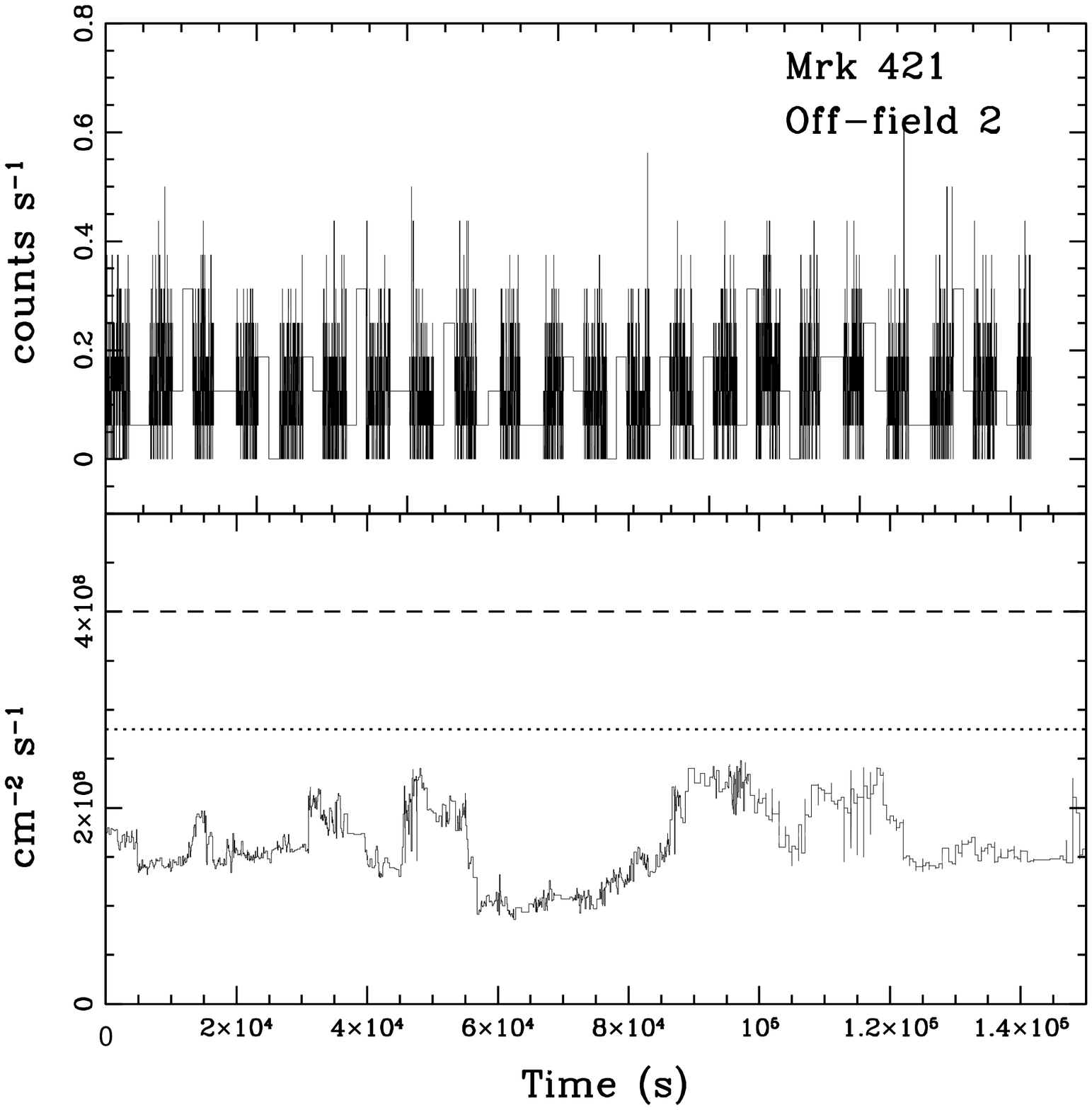}
\end{center}
\caption{XIS1 0.5-2.0 keV light curve ({\it upper panel}) and 
solar wind proton flux data from OMNIWeb ({\it lower panel}) in 
Mrk~421 Off-field1 (top) and 
Off-field2 (bottom) observation periods. The {\it dashed} and {\it dotted} 
lines show the proton flux threshold level 
($4 \times 10^{8}~$cm$^{-2}$~s$^{-1}$) used in this 
investigation and the average proton flux at 1~AU 
($2.8 \times 10^{8}~$cm$^{-2}$~s$^{-1}$) respectively.
}
\label{fig3}
\end{figure}

\clearpage

\begin{figure}
\begin{center}
\includegraphics[width=10cm,height=10cm]{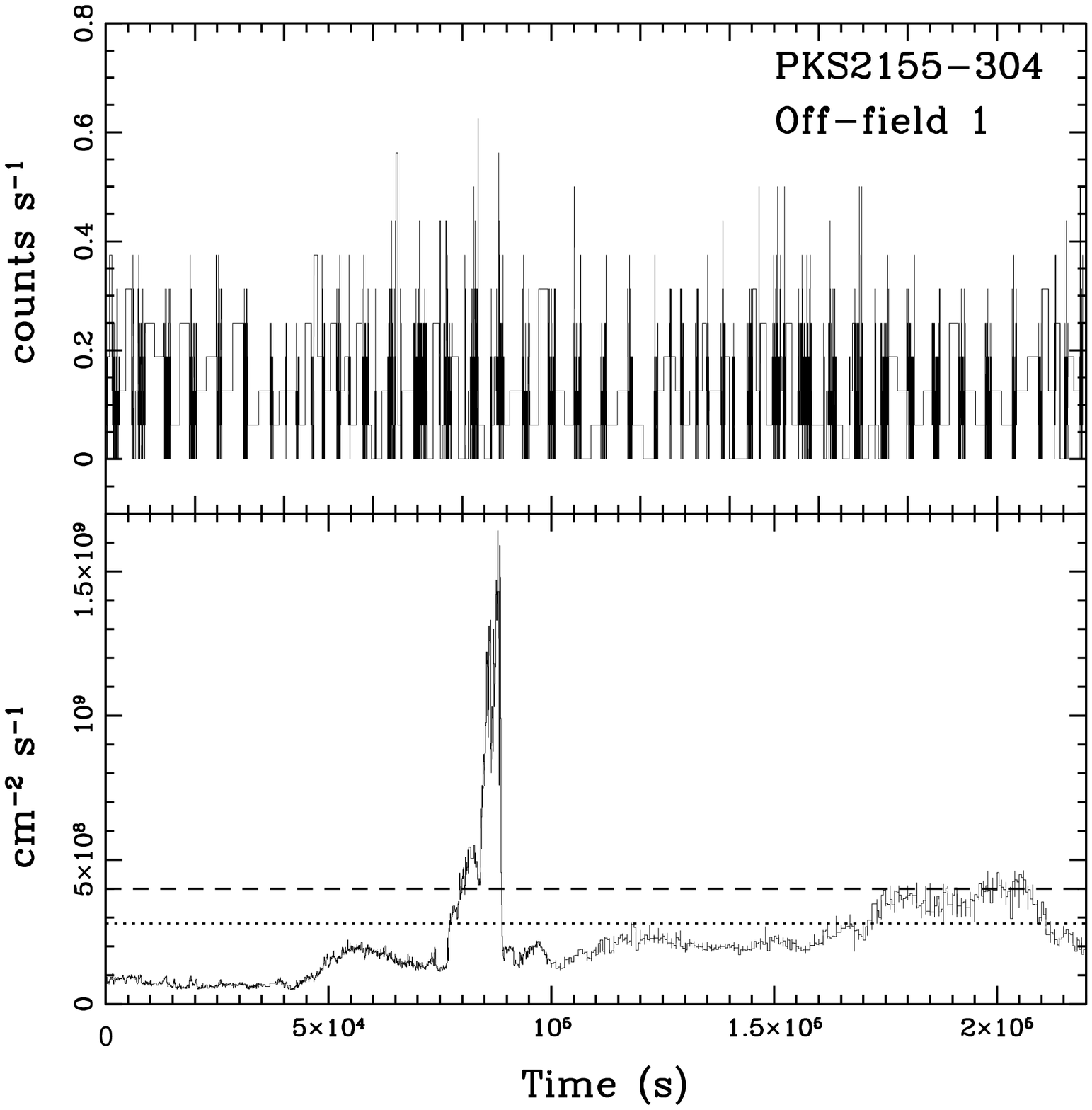}
\hspace{1cm}
\includegraphics[width=10cm,height=10cm]{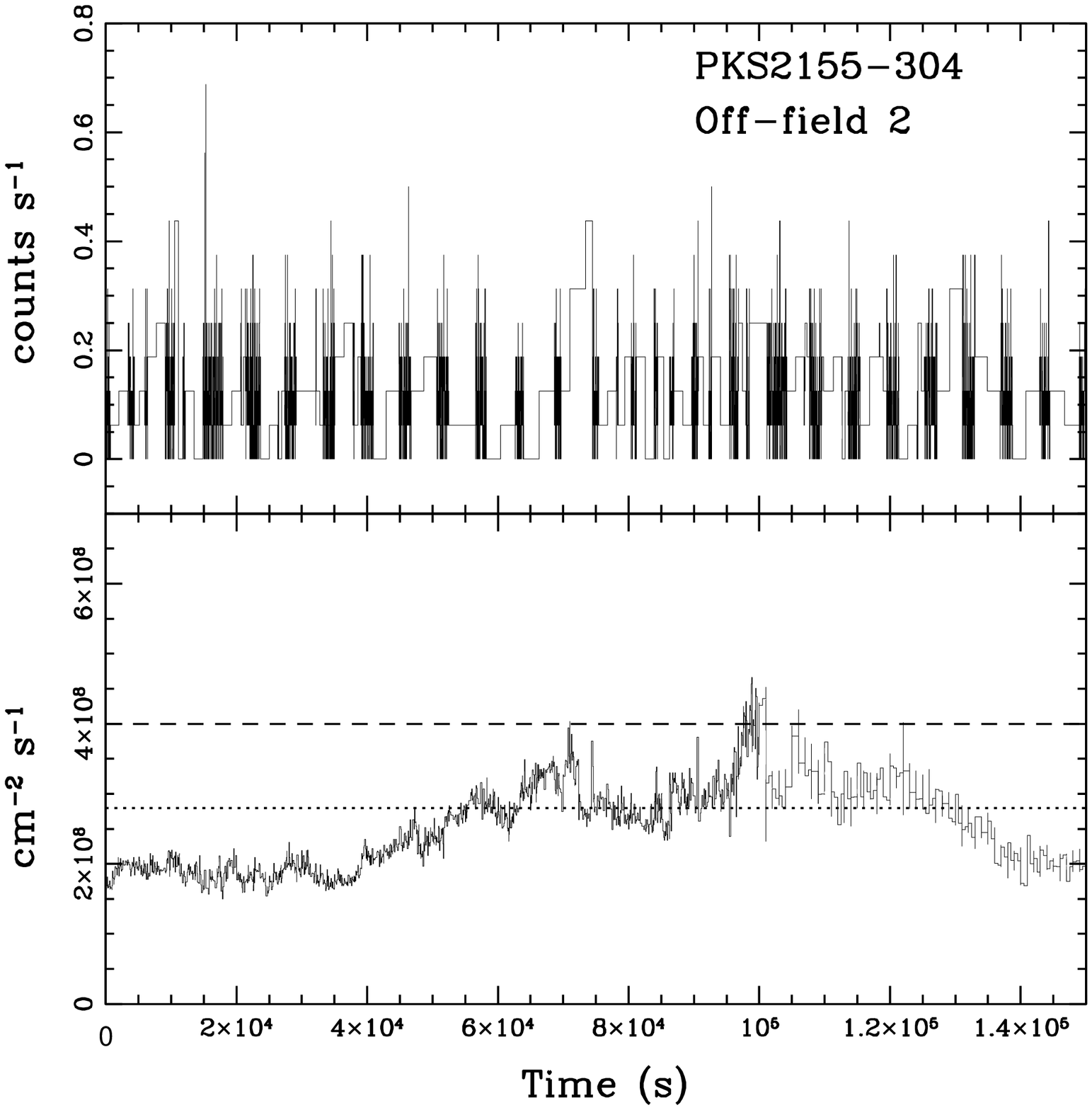}
\end{center}
\caption{Same as fig. 3 for PKS~2155-304.}
\label{fig4}
\end{figure}

\clearpage

\begin{figure}
\begin{center}
\includegraphics[width=10cm,height=10cm]{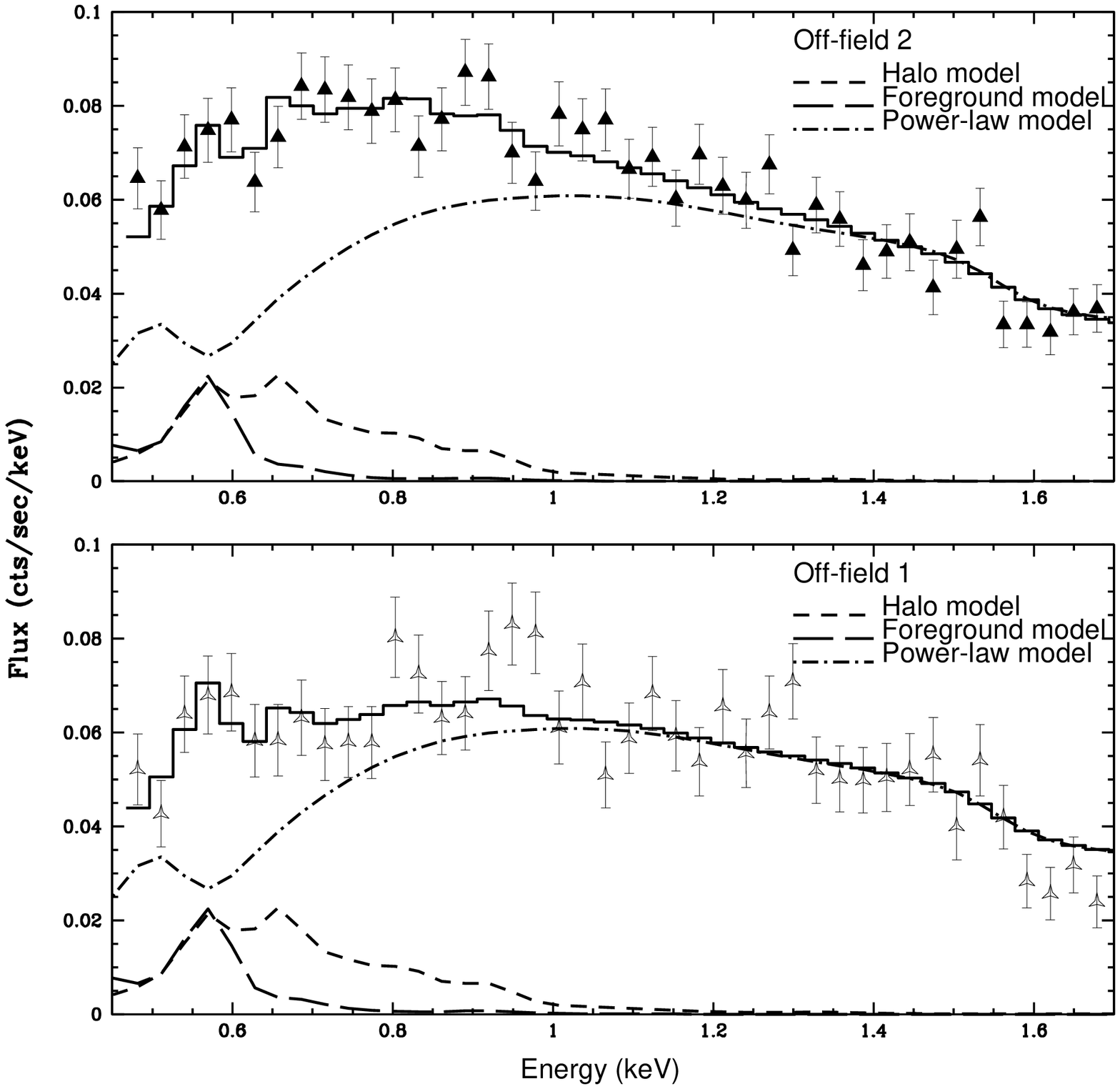}
\hspace{1cm}
\includegraphics[width=10cm,height=10cm]{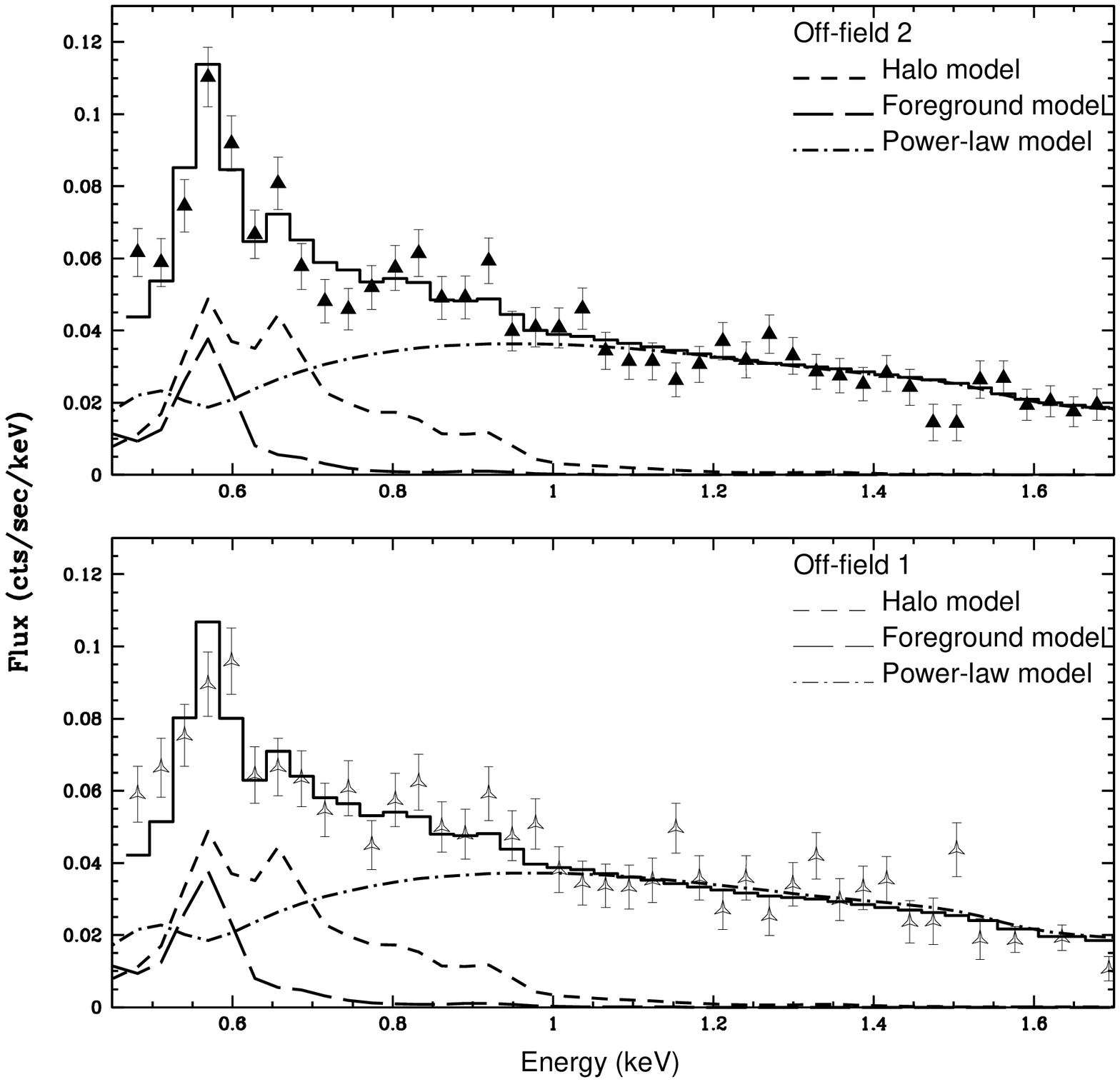}
\end{center}
\caption{Simultaneous fits to Off-field~1 and Off-field~2 Suzaku spectra 
of regions near Mrk~421({\it top}) and PKS2155-304({\it bottom}). The dashed 
curves represent the galactic halo emission model.
}
\label{fig5}
\end{figure}

\clearpage

\begin{figure}
\begin{center}
\includegraphics[width=10cm,height=10cm]{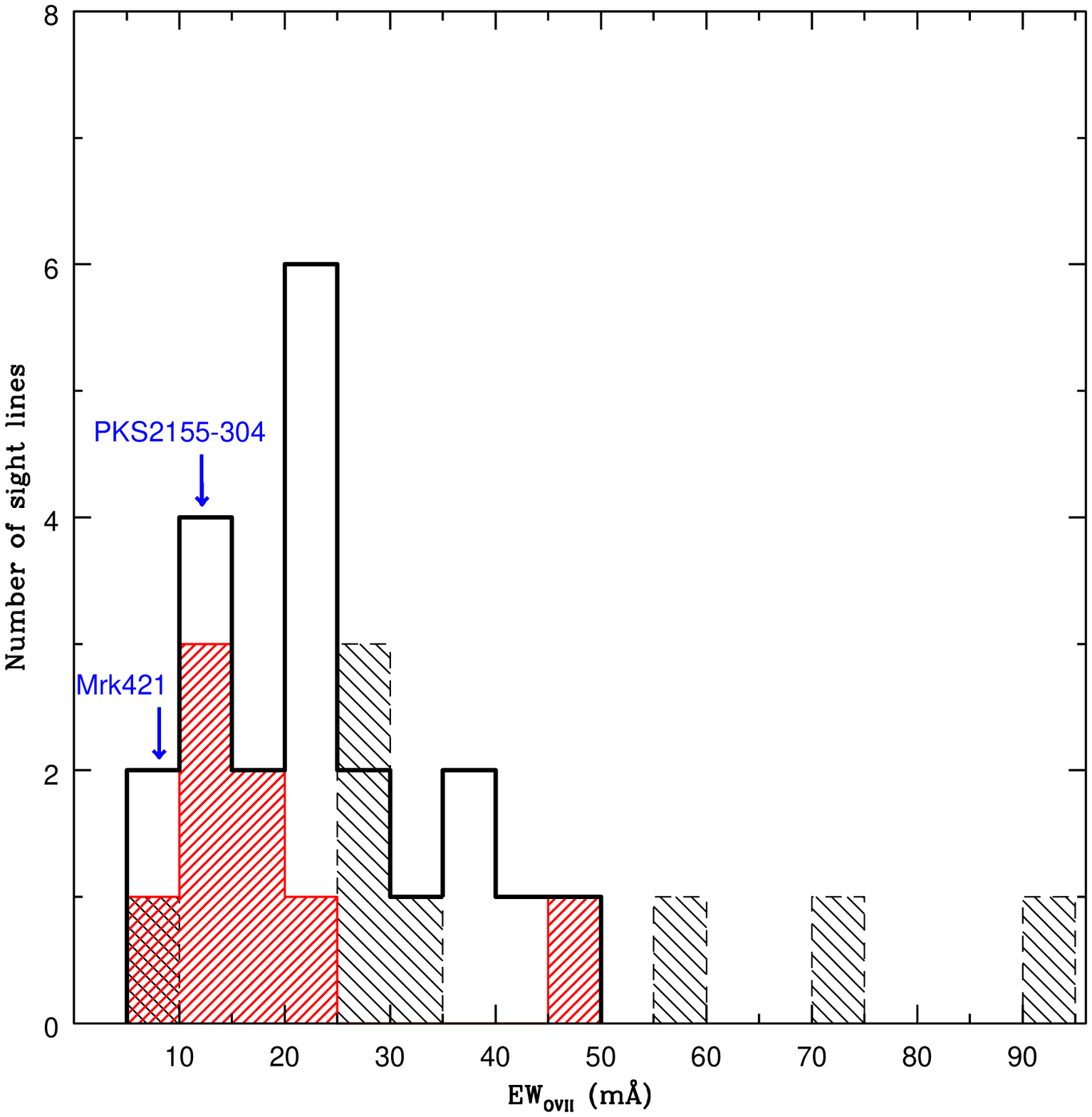}
\hspace{1cm}
\includegraphics[width=10cm,height=10cm]{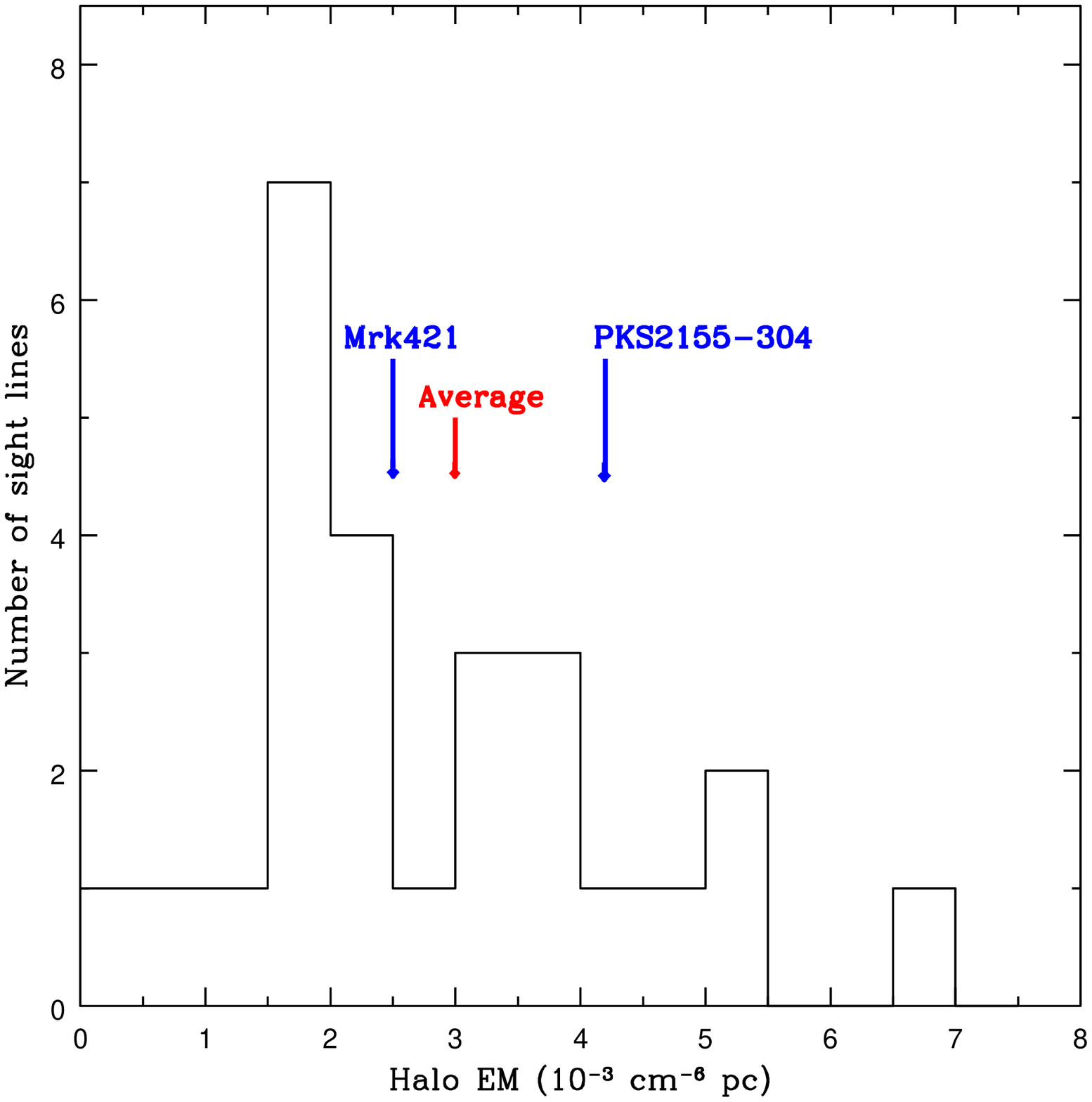}
\end{center}
\caption{{\it Top}: Distribution of \ovii$_{\rm K\alpha}$ line EW for the 
parent sample (Paper-I). The red shaded region corresponds to the sub-sample 
used in Paper-I. {\it Bottom}: Distribution of galactic halo emission 
measure using data from Henley et al. (2010). The vertical arrows mark the 
\ovii$_{\rm K\alpha}$ EW and emission measure towards Mrk~421 and 
PKS2155-304. }
\label{fig6}
\end{figure}

\clearpage

\begin{figure}
\begin{center}
\includegraphics[width=10cm,height=10cm]{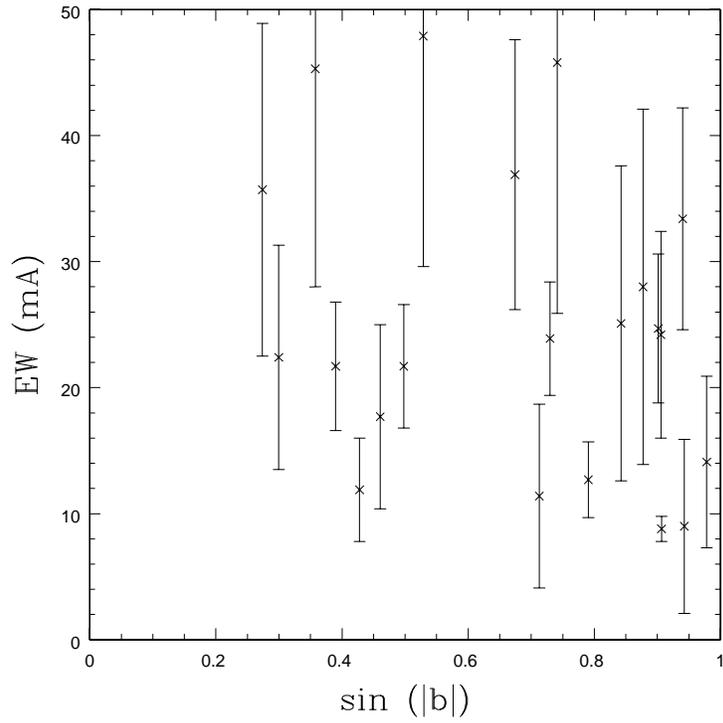}
\end{center}
\caption{
Equivalent widths of \ovii absorption lines are plotted vs. $\sin(|b|)$
where $b$ is the Galactic latitude of the sightline. The data are from
Paper-I (their figure 3). We see no anticorrelation between EW and
$\sin(|b|)$; this implies that the Galactic thick disk does not
contribute significantly to the observed strengths of absorption lines.
}
\label{fig6}
\end{figure}

\clearpage

\begin{deluxetable}{lcccc}
\tablewidth{0pc}
\tablecaption{Details of our Suzaku observations.}
\tablehead{
\colhead{Target} & \colhead{Observation ID} & \colhead{Start Time} & 
\colhead{End Time} & \colhead{Exposure\tablenotemark{a}}\\
\colhead{} & \colhead{} & \colhead{UT} & \colhead{UT} & \colhead{ks}\\
}
\startdata
\textbf{Mrk421-off-field1} & 503082010  &  2008-04-29~18:32:39  &  2008-05-02~08:30:08 & 48 \\
\textbf{Mrk421-off-field2} & 503083010  &  2008-05-02~08:31:41  &  2008-05-04~17:30:19 & 50\\
\textbf{PKS2155-off-field1}& 504086010  &  2009-11-09~01:34:20  &  2009-11-10~19:25:07 & 45\\
\textbf{PKS2155-off-field2}& 504087010  &  2009-11-11~10:45:08  &  2009-11-13~07:37:24 & 72\\
\enddata
\tablenotetext{a}{Good-Time-Interval after removing the exposure 
affected by high proton flux.}
\end{deluxetable}

\clearpage

\clearpage

\begin{deluxetable}{lcccccccc}
\rotate
\tablecolumns{10}
\tablewidth{0pc}
\tablecaption{Model parameters of the spectral fits}
\tablehead{
\colhead{Dataset(s)}  & \multicolumn{2}{c}{Local component\tablenotemark{a}} & \multicolumn{2}{c}{Galactic Halo} & \colhead{} & \multicolumn{2}{c}{Power Law}  & \colhead{\it {$\chi^2/dof$}}\\
\cline{2-3} \cline{4-5} \cline{7-8}\\
\colhead{} & \colhead{$\log T$} & \colhead{E.M.\tablenotemark{b}}  & \colhead{$\log T$} & \colhead{E.M.} & \colhead{} & \colhead{$\Gamma\tablenotemark{c}$} & \colhead{Norm\tablenotemark{d}} & \colhead{}\\
\colhead{} & \colhead{K} & \colhead{$10^{-3}~cm^{-6}~pc$} & 
\colhead{K} & \colhead{$10^{-3}~cm^{-6}~pc$} & \colhead{} & \colhead{} & \colhead{} & \colhead{}}
\startdata
\textbf{Mrk~~421} & 6.08 & 7.3 &  $6.32\pm0.03$  & $2.5\pm0.6$ & &  &   & 289/244\\
Off-field~1 &  &    &  &  &  & $1.54\pm0.04$  & $17.3\pm0.6$   & \\
Off-field~2 &  &    &  &  &  & $1.76\pm0.04$  & $18.3\pm0.5$   & \\
\textbf{PKS2155-304} & 6.08 & 9.7  & $6.36\pm0.02$  & $4.2\pm0.8$  & &  & & 313/281\\
Off-field~1 &  &  &  &  &  & $1.60\pm0.05$  & $9.3\pm0.3$     & \\
Off-field~2 &  &  &  &  &  & $1.65\pm0.04$  & $9.5\pm0.3$     & \\
\enddata
\tablenotetext{a}{Parameters of this component kept frozen at values as described 
in the text.}
\tablenotetext{b}{Emission Measure}
\tablenotetext{c}{Index of absorbed power law fit}
\tablenotetext{d}{Normalization of power law fit at 1~keV in units of 
$\textrm{photons~keV}^{-1}~\textrm{s}^{-1}~\textrm{cm}^{-2}~\textrm{sr}^{-1}$ }
\end{deluxetable}

\clearpage

\begin{deluxetable}{lcc}
\tablewidth{0pc}
\tablecaption{$O_{VII}$ and $O_{VIII}$ line intensities}
\tablehead{
\colhead{Target} & \colhead{$O_{VII}$} & \colhead{$O_{VIII}~K\alpha
+O_{VII}~K\beta$}\\
\colhead{} & \colhead{ph~s$^{-1}~$cm$^{-2}~$sr$^{-1}$} & 
\colhead{ph~s$^{-1}~$cm$^{-2}~$sr$^{-1}$} \\
}
\startdata
\textbf{Mrk421-off-field1}  & $5.5\pm0.9$ &  $1.4\pm0.5$ \\
\textbf{Mrk421-off-field2}  & $5.6\pm0.7$ &  $1.8\pm0.3$ \\
\textbf{PKS2155-off-field1} & $7.4\pm0.7$ &  $1.9\pm0.4$ \\
\textbf{PKS2155-off-field2} & $8.3\pm0.7$ &  $2.6\pm0.4$ \\
\enddata
\end{deluxetable}

\appendix

\newpage
\section{Appendix}

Here we show explicitly how a two component density distribution affects
the resultant halo path-length and mass.

 Let us assume that there is a high density component with $n(High)
=1\times 10^{-3}$ cm$^{-3}$ and pathlength L(High)=1kpc. We can identify
this component with the Galactic disk, but that is not important.  Let
us also say there is a halo of density $n(Low)=1\times 10^{-5}$
cm$^{-3}$.

The column density of the high density component is then $N_H=3\times
10^{18}$ cm$^{-2}$ and the emission measure EM(High)$=n(High)^2L(High)=
3\times 10^{15}$ in units of cm$^{-5}$.

The EM would be strongly biased by the high density component, so let us
assume it contributes $90\%$ to the total emission measure. Therefore
the total EM$=3.3\times 10^{15}$ in the same units.

By construction, the EM from the low density component is only $10\%$,
that is EM(Low)$=3.3\times 10^{14}$ in the same units. Given this
EM(Low) and n(Low) we can now calculate the path-length of the low
density component: L(Low)$=300$ kpc. And its column density
N$_H(Low)=3\times 10^{19}$cm$^{-2}$.

The total disk $+$ halo column density is then N$_H=3.3\times
10^{19}$cm$^{-2}$. Thus we have completely defined a two component
medium, with the high density component (disk) dominating the observed
EM while the low density component (halo) dominating the observed column
density. The EM we measure is EM$=3.3\times 10^{15}$ cm$^{-5}$ and the
total column density we measure is N$_H=3.3\times 10^{19}$cm$^{-2}$.

If we now make the assumption of a single density medium, and derive
density and pathlength from the total observed EM and N$_H$, the values
we get are as follows. The average density is $n=EM/N_H=1\times 10^{-4}$
cm$^{-3}$ and the average pathlength L$=N_H/n=110$ kpc. Thus, under the
assumption of a single density plasma, the derived density is an order of
magnitude higher and the derived pathlength is about a factor of three
lower than the actual density and pathlength of the halo.

The actual mass of the halo (low density component) is
M(Low)$=n(Low)L(Low)^3=7.3\times 10^{66}$ in dimensionless units
(ignoring constants). The derived mass under the single component
assumption, however, is M$=nL^3=3.6\times10^{66}$ in the same
dimensionless units. Thus the mass derived assuming a single density
component is a factor of two lower than the actual mass.

Thus, the assumption of a single density halo underpredicts the halo
pathlength and mass. We have explicitly presented a two component system
here for simplicity, but it shows that {\it any} density gradient would
similarly strengthen our results. In this example, we have considered
that the high density component contributes $90\%$ to the total
EM. Higher contribution to the EM would make the halo even more
extended. We thus conclude that any disk contribution, clumping, and/or
a declining density profile would make the halo more extended and more
massive, thus strengthening our result.

\end{document}